\documentclass{article}
\usepackage{multirow}
\usepackage{microtype}
\usepackage{graphicx}
\usepackage{subfigure}
\usepackage{booktabs} 

\usepackage{hyperref}

 \usepackage[accepted]{icml2024}

\usepackage{amsmath}
\usepackage{amssymb}
\usepackage{mathtools}
\usepackage{amsthm}

\usepackage[capitalize,noabbrev]{cleveref}

\theoremstyle{plain}

\theoremstyle{definition}

\theoremstyle{remark}

\usepackage[textsize=tiny]{todonotes}

\usepackage[marginal]{footmisc}
\icmltitlerunning{DTAMLP: Denoise Time-aware MLP for Session-based Recommendation}

\begin{document}

\twocolumn[
\icmltitle{DTAMLP: Denoise Time-aware MLP for Session-based  Recommendation}



\icmlsetsymbol{equal}{*}

\begin{icmlauthorlist}
\icmlauthor{Jiamu Zheng}{to,equal}
\icmlauthor{Xiaojun Shan}{to}

\end{icmlauthorlist}

\begin{center}
\normalsize \textsuperscript{1}University of Electronic Sciences and Technology of China
\end{center}

\icmlaffiliation{to}{University of Electronic Sciences and Technology of China}



\vskip 0.3in
]

\footnote{* Carried out by Jiamu during a research internship at Prof. Bo Yang's lab, School of Computer Science and Engineering, UESTC (thanks to Prof. Yang and the lab's senior members for their advice). Completed in May 2023 but not released publicly until now due to unavoidable circumstances. As Jiamu's first independent research work, it marks the start of his research journey; though its novelty falls short of a top-tier paper, we believe the insights are worth recording, and share it here as a memento.\\}


\begin{abstract}
  This paper reports two empirical findings on session-based recommendation (SBR), unified in a single model, DTAMLP. First, existing time-aware and GNN-based models (e.g., TiSASRec, SR-GNN) treat every click-time interval as equally informative, even though very short dwell times often reflect accidental clicks carrying little preference signal -- a phenomenon we call \textit{sporadic noise}. We show that a lightweight, plug-and-play \textit{weight fusion} module, blending a model's attention weight with a threshold-capped time-interval weight, can be inserted into such models with almost no architectural change and yields a consistent accuracy gain; we view this as the most directly verifiable contribution of this work. Second, we revisit an under-explained observation from FMLP-Rec, where a learnable frequency-domain filter on item embeddings improves accuracy, and offer a possible explanation: time-domain behavior mixes several entangled psychological preferences, and a frequency-domain view may let a model separate and down-weight such \textit{preference noise} more naturally -- an interpretive conjecture rather than a proven mechanism. Building on both insights, DTAMLP, an all-MLP framework combining weight fusion and FFT-based filtering, is validated on Diginetica and RetailRocket. While this system-level design reflects the state of the field circa 2023 rather than a state-of-the-art claim, ablations confirm the two mechanisms contribute complementary, non-redundant improvements.
\end{abstract}
\section{INTRODUCTION}
\label{1-intro}

Session-based recommendation(SBR) is a critical research area that has found broad applications in various domains such as e-commerce, short video recommendation, and media streaming platforms \cite{koren2009matrix,sarwar2001item,shani2005mdp,zhang2019deep,wang2021survey}. Unlike conventional recommendation systems that rely on user identity and historical activities, session-based recommendation systems primarily focus on anonymous user actions. Given the prevalence of anonymous interactions on the Internet, it is crucial to model the constrained behavior exhibited within a given session and generate recommendations based on this information. In contrast, conventional recommendation methods that rely on sufficient user-item interactions encounter difficulties in producing precise results under such circumstances.

To better model the short-term behavioral preferences of users and ensure an accurate representation of items, early effort \cite{hidasi2016gru4rec,li2017narm,sitoula2019context} has been made relying on Convolutional Neural Network (CNN) and Recurrent Neural Network (RNN) architectures. Furthermore, some works use more advanced architectures such as attention mechanism \cite{bian2021contrastive,li2017neural,he2021locally}, and Graph Neural Networks (GNN) \cite{chen2020lessr,qiu2019rethinking,wu2019srgnn}. Recently, there has been a surge in the popularity of Transformer-based models and Multi-Layer Perceptron(MLP)-based  architectures \cite{li2020time,sun2019bert4rec}, which have achieved remarkable results in this context.

This paper originates from a simple empirical observation made while we were experimenting with time-aware SBR models, rather than from a top-down design of a new architecture. We report this observation, together with a second, related observation on frequency-domain filtering, and finally describe how we combined the two into a single system, DTAMLP, for the sake of completeness.

\textbf{\textit{Observation 1: sporadic noise in click-time intervals.}} Most time-aware models \cite{li2020time} take the time interval between consecutive clicks as an input feature and let an attention mechanism decide how much each item should contribute to the session representation, implicitly assuming that a longer dwell time always means stronger interest and treating every interval as an equally reliable signal. This assumption breaks down in a common daily scenario: when shopping on an e-commerce platform, users may come across a picture or a description that is eye-catching but unclear about the product, which may suddenly pique their interest and cause them to click on it. However, upon entering the product page and learning about the product, users may realize that they do not need it and exit the page in a short time. Such clicks are only weakly related to actual preference and, if treated the same as other clicks, can distort the learned session representation. We refer to this as \textbf{\textit{sporadic noise}}. What makes this observation practically useful is that a very small, plug-and-play intervention -- fusing the model's existing attention weight with a simple, threshold-capped function of the time interval -- is enough to noticeably improve accuracy on off-the-shelf models such as TiSASRec and SR-GNN, \emph{without} modifying their backbone architecture (Table~\ref{table:ti}). We see this as the most directly verifiable finding of this paper.

\textbf{\textit{Observation 2: a possible explanation for frequency-domain filtering.}} Independently of the above, we noticed that FMLP-Rec \cite{zhou2022fmlp} obtains a solid improvement by applying a learnable filter to item embeddings in the frequency domain, yet the original work does not explain \emph{why} such filtering should help a recommendation task where the input is not literally a physical signal. Existing SBR methods \cite{hidasi2016gru4rec,li2017narm, ren2019repeatnet} model user preferences purely from features in the time domain, which are an explicit, entangled mixture of a user's underlying (and possibly concurrent) psychological preferences: when choosing the next item, some of these underlying preferences are relevant to the decision, while others contribute little and effectively act as noise. Our conjecture is that frequency-domain filtering is effective precisely because it gives the model an alternative basis in which such entangled, weakly-relevant preference components -- which we refer to as \textbf{\textit{preference noise}} -- are easier to separate and down-weight than in the raw time domain. We want to be explicit that this is an interpretive hypothesis, illustrated with intuition and a qualitative example (Figure~\ref{fig:fourier}) rather than a rigorously proven causal mechanism.

Apart from these two observations, we also noticed that most SBR works rely heavily on nonlinear operations (such as RNN and GNN), which can leave session embeddings and item embeddings in inconsistent representation spaces. When this happens, similarity-based prediction \cite{hidasi2016gru4rec,li2017narm} loses part of its theoretical justification. To keep our system-level implementation clean, we adopt an existing representation-consistent design \cite{hou2022core} for this part rather than treating it as a new contribution of this paper.

To examine whether the two observations above could be combined into a coherent, self-contained model -- which was also a natural goal to pursue at the time (2023) -- we assembled the novel \textbf{D}enoise \textbf{T}ime-\textbf{A}ware \textbf{MLP} (DTAMLP) framework. We model the session embedding as a linear combination of item embeddings in the same representation space as the items themselves, use an FFT-Transformer module to reduce preference noise when computing each item's contribution weight, and use a weight fusion module to reduce sporadic noise via time intervals. We view DTAMLP mainly as an integration vehicle that lets us check, through ablation, whether the two mechanisms above are complementary rather than redundant; the system-level accuracy numbers reflect the state of the field circa 2023 and are reported here mainly as supporting evidence rather than as a state-of-the-art claim.

We summarize the content of this paper as follows: (1) We report a simple, low-cost weight fusion intervention that can be plugged into existing time-aware or attention-based SBR models with almost no architectural change, and show that it yields a consistent accuracy improvement on TiSASRec and SR-GNN -- we consider this the central, most directly verifiable finding of the paper. (2) We revisit the under-explained effectiveness of frequency-domain filtering in FMLP-Rec and offer a conjecture -- namely, the notion of preference noise -- for why separating preferences in the frequency domain may be easier than doing so in the time domain; this explanation is offered as an interpretive hypothesis rather than a proven result. (3) We integrate both observations, together with a representation-consistent embedding design adapted from prior work, into a single all-MLP model, DTAMLP, and report its performance on Diginetica and RetailRocket together with ablation studies showing that the two mechanisms contribute complementary improvements.

\section{RELATED WORK}
\label{2-related-work}

\textit{Traditional SBR} mainly relies on Collaborative Filtering (CF) techniques \cite{koren2009matrix}, item-based methods \cite{sarwar2001item}, and Markov chains \cite{shani2005mdp}. These methods largely treat session items as independent entities or rely on low-order transition statistics, and generally do not model the fine-grained sequential and temporal structure within a session.

Recently, Deep Learning (DL) technology has been widely used in recommender systems \cite{zhang2019deep}. Hidasi et al. \cite{hidasi2015session} take the lead in applying RNN for SBR. They use GRU to predict the next action and obtain good performance. Tan, Xu, Liu \cite{tan2016improved} further propose data augmentation and avoiding distribution skew of input data, in RNN-based SBR to improve the recommendation performance. Then a surge of works leverages other neural network architectures for a sequential recommendation, e.g., CNN \cite{tang2018personalized}, GNN \cite{tan2021sparse,xu2019graph}, and Transformer \cite{kang2018sasrec, sun2019bert4rec}. Based on these neural network architectures, various studies introduce other contextual information (e.g., item attributes and reviews) by adding memory networks \cite{huang2018improving}, hierarchical structures \cite{li2019review}, data augmentation \cite{wang2021counterfactual,zhang2021causerec}, and pre-training techniques \cite{bian2021contrastive,bian2021novel,zhou2020s3}.  \textit{GNN-based SBR}\cite{wu2019srgnn}\cite{xu2019gcsan}\cite{pan2021graph} has gained popularity with the booming of deep learning and GNN, which is a connection model that leverages message passing between graph nodes to extract the dependency of graphs. 

Multi-Layer Perceptron (MLP) \cite{yair1988boltzmann} is a classical feed-forward neural network that consists of fully-connected neurons and non-linear activation functions. Recently, many studies doubt the necessity of CNN \cite{tolstikhin2021mlp} and self-attention network \cite{li2021localvit,lian2021mlp} and propose to use MLP to replace these architectures. GFNet \cite{rao2021global} adopts 2D Fourier transform with the global filter layer to learn long-term spatial dependencies in the frequency domain and FMLP \cite{zhou2022fmlp} uses 1D Fourier transform to filter the noise in item embedding, which shows competitive performance to Transformer-based models.

Nowadays, some models incorporate time information directly into the model, which we refer to as \textit{Time-aware SBR}. TiSASRec \cite{li2020time} utilizes a time-aware self-attention framework that can effectively capture the absolute positions of session items and the time intervals between consecutive items in the session. TNARM \cite{wang2022session} uses a time-aware neural attention network to model dynamic user preferences. To the best of our knowledge, these models treat every observed time interval as an equally reliable signal of user interest; Section~\ref{3-Analysis} presents our empirical observation that this assumption can be violated by what we call sporadic noise, and that a simple correction for it improves two representative models from this line of work.

\section{PRELIMINARIES}
\label{3-Analysis}
\subsection{Promblem Formulation}
Session-Based Recommendation (SBR) is a critical task that involves predicting the next item that a user is likely to click based on their present behavior sequence. In this paper, we present a novel DTAMLP designed specifically for the SBR task. To provide an in-depth understanding of DTAMLP, we will offer a precise definition of SBR first. 

Given a session $s=\{(v_1,t_1), (v_2,t_2), \cdots, (v_n, t_n)\}$ consisting of $n\ll m$ items, where each item is represented by a serial number $v_i$, and click timestamp $t_i$. Additionally, $m$ denotes the total number of items and $n$ denotes that there are n items in session $s$. The primary aim of DTAMLP is to generate an output sequence $y$ which is a sorted list of potential items that the user may click next. Each item in the list denoted as $y_j$ (1 $\leq j \leq m$), represents the probability of the user clicking that particular item. Since recommendation systems often suggest multiple items simultaneously, we propose that the top-k (1$ \leq k \leq m$) items in $y$ can be recommended to the user.
\begin{figure}[h]
    \centering
 \includegraphics[width=0.45\textwidth]{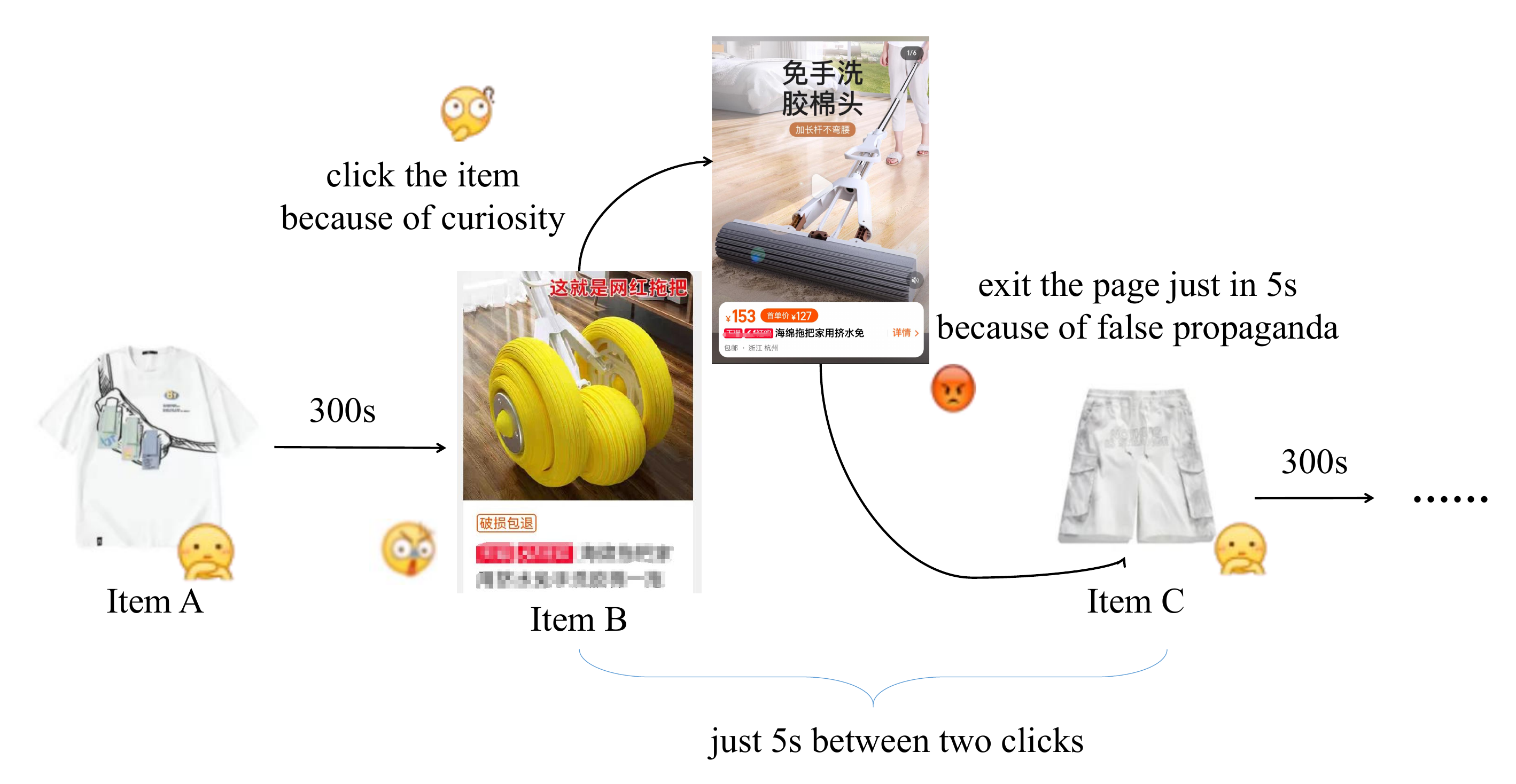}
  \caption{An example of shopping on e-commerce platforms. (The time above means the time interval between two clicks.) Sometimes you may click "Item B" with curiosity because its advertising image is really attractive. However, after you click the item and know what it is, you may exit the page and click another item C in a short time because you don't need it or feel deceived by its false propaganda. }

  \label{fig:example}
 \end{figure}

\subsection{Analysis on Two Types of Noise}
In this part, we introduce Sporadic Noise and Preference Noise. The former is supported by a direct, plug-and-play verification experiment on off-the-shelf models (Table~\ref{table:ti}), which we consider the core empirical finding of this paper. The latter is presented as an interpretive conjecture that motivated our design choice, rather than as an independently proven phenomenon.
\subsubsection{Sporadic Noise and Time Interval}
\label{3.2.1}
Assuming that during session $s$, a user $i$ has interacted with three items, $Item_A$, $Item_B$, and $Item_C$. Based on existing interaction records, we aim to predict the next potential item of interest, $Item_D$. Heuristically, the browsing duration of an item can serve as an indicator of the user's interest in the item. However, existing datasets only provide click timestamps and lack information regarding the end of browsing. Accordingly, we approximate the browsing duration using the time interval between two adjacent clicks which we define as "\textit{interval}". This approximation serves as a proxy for the user's attention toward the item.  For instance, let us denote the \textit{interval} of $Item_A$ as 300 seconds, which means that the user clicks $Item_B$ 300 seconds after clicking $Item_A$. In the given sequence of interaction records, the $Item_A$'s \textit{interval} is observed to be significantly long, suggesting that the user might have spent a considerable amount of time browsing $Item_A$. However, it is also possible that the user may have temporarily suspended the browsing session to attend to other tasks. Hence, we define a threshold $\eta$ as the upper limit of the \textit{interval}. Intuitively, when a user spends a shorter duration browsing certain items' pages, it implies that the user maybe has a lesser interest in those items. We classify such confusing items as \textbf{sporadic noise}. 
Since the user may have limited interest in such items, we need to weaken their impact on modeling user preference. Besides, since the user has clicked the item, it shows that the item still has some attractive elements. Accordingly, we should retain a portion of their impact rather than simply leave it out of consideration
. Herein, we introduce a novel weight fusion module to reduce its impact on modeling user preference. For instance, the interval between $Item_B$ and $Item_C$ is only 5 seconds. Given the time required for user action and reaction, it's safe to assume that the user spends less than 5 seconds browsing  $Item_B$, which is a very short browsing duration. Therefore, we can infer that $Item_B$ may be sporadic noise. When modeling user preferences in subsequent stages, we should reduce the impact of $Item_B$. Some time-aware models usually use attention mechanisms that will generate a weight $\alpha_1$ for $Item_B$. 

\begin{equation}
\label{attention}
\alpha_1 = AttentionLayer() 
\end{equation}

Therefore, we can slightly reduce the weight of $Item_B$ in an attention mechanism. To do so, We apply the softmax operation to the intervals with threshold $\eta$. Accordingly, $Item_B$ will obtain weight $\alpha_2$, calculated as follows:

\begin{equation}
\label{softmax}
\alpha_2 = softmax(min(interval, \eta))
\end{equation}


Our model utilizes an approach similar to linear interpolation to merge two weights, and we define $\beta$ as the blend coefficient. The resulting impact weight $\alpha$ is calculated by:
\begin{equation}
\label{weight}
\alpha = (1-\beta)\times \alpha_1 + \beta \times \alpha_2
\end{equation}

The above steps illustrate how we reduce the impact of sporadic noise using click time intervals with $\eta$. Since the time intervals are capped, after applying the softmax operation and merging the weights, items with larger interval values will still have significant weights. On the contrary, for items with smaller interval values, we intentionally decrease their final weight to reduce the potential effects of sporadic noise in our modeling process.

We deliberately keep this intervention as simple as possible so that it can be verified as a plug-and-play addition, without touching the backbone of an existing model. We first apply it to \textbf{TiSASRec} on MovieLens-1m: with all interval values sorted in ascending order, we set $\eta$ as the 20$th$ percentile and $\beta$ as a fixed value in $\{0.1, 0.2, 0.3\}$, and observe a consistent improvement over the original TiSASRec across all tested $\beta$ values (Table~\ref{table:ti}, ``Origin'' denotes the unmodified backbone). We then apply the same idea, with $\beta$ set as a learnable parameter instead of a fixed constant (``Learnable Para''), to \textbf{SR-GNN} \cite{wu2019srgnn} on Diginetica -- a GNN-based model that has no explicit time-awareness mechanism to begin with -- and again observe a clear gain over the original SR-GNN (Table~\ref{table:ti}). Since TiSASRec and SR-GNN are verified on two different datasets, with metrics appropriate to each, we report each backbone in its own block of Table~\ref{table:ti} rather than in a single dataset-by-dataset grid, so that every reported cell corresponds to an actual run. We consider this pair of results, obtained on two architecturally very different backbones (a self-attention model and a graph neural network) with an almost cost-free modification, to be the most direct and reproducible piece of evidence in this paper -- and the core empirical evidence for Insight 1 (sporadic noise): click-time intervals contain exploitable sporadic-noise signal that current time-aware and GNN-based SBR models do not fully utilize.

\begin{table}[!ht]
    \small
    \centering
    \caption{Plug-and-play verification of the Weight Fusion module, without any other change to the original backbone.}
    \label{table:ti}
    \setlength{\abovecaptionskip}{1pt}
    \setlength{\belowcaptionskip}{1pt}
    \begin{tabular}{llcc}
        \toprule
        \multicolumn{4}{c}{\textbf{TiSASRec on MovieLens-1m}} \\
        \midrule
        \textbf{Variant} & \textbf{Setting} & \textbf{NDCG@10} & \textbf{HR@10} \\
        \midrule
        Origin & -- & 56.8163 & 79.3377 \\
        + Weight Fusion & $\beta = 0.1$ & 57.1112 & 80.1324 \\
        + Weight Fusion & $\beta = 0.2$ & 57.2038 & 80.1158 \\
        + Weight Fusion & $\beta = 0.3$ & \textbf{57.4986} & \textbf{80.1490} \\
        \midrule
        \multicolumn{4}{c}{\textbf{SR-GNN on Diginetica}} \\
        \midrule
        \textbf{Variant} & \textbf{Setting} & \textbf{MRR@20} & \textbf{Recall@20} \\
        \midrule
        Origin & -- & 15.8268 & 47.4723 \\
        + Weight Fusion & Learnable $\beta$ & \textbf{18.0137} & \textbf{48.2895} \\
        \bottomrule
    \end{tabular}
\end{table}

\subsubsection{Preference Noise: A Conjecture Motivated by Frequency-Domain Filtering}
We now turn to a second, more speculative observation. Most SBR approaches model user preferences based solely on temporal features, such as the order and timing of interaction records. However, behavior observed in the time domain is the outcome of a user's underlying preferences, and a user may hold multiple, possibly conflicting psychological preferences at the same time; only some of them are relevant to the item eventually clicked. If this is the case, denoising strategies that operate purely on the time domain -- such as the weight fusion mechanism above -- may be structurally unable to separate these entangled preference components, because they never leave the time domain in the first place.

This line of reasoning was prompted by an observation in \cite{zhou2022fmlp}: applying a learnable Fourier filter to item embeddings was reported to improve recommendation accuracy, but the original work does not explain \emph{why} this should be the case for embeddings that are not literally a physical signal. In what follows, we lay out our conjecture for why a frequency-domain view may make such preference components easier to separate, which we refer to as \textbf{Preference Noise}. We want to be explicit that this is an interpretive hypothesis intended to motivate our design choice (Section~\ref{4.3}), not a phenomenon we independently prove in this paper. We start with a brief review of the Fourier transform itself.

\begin{figure}
    \centering
 \includegraphics[width=0.45\textwidth]{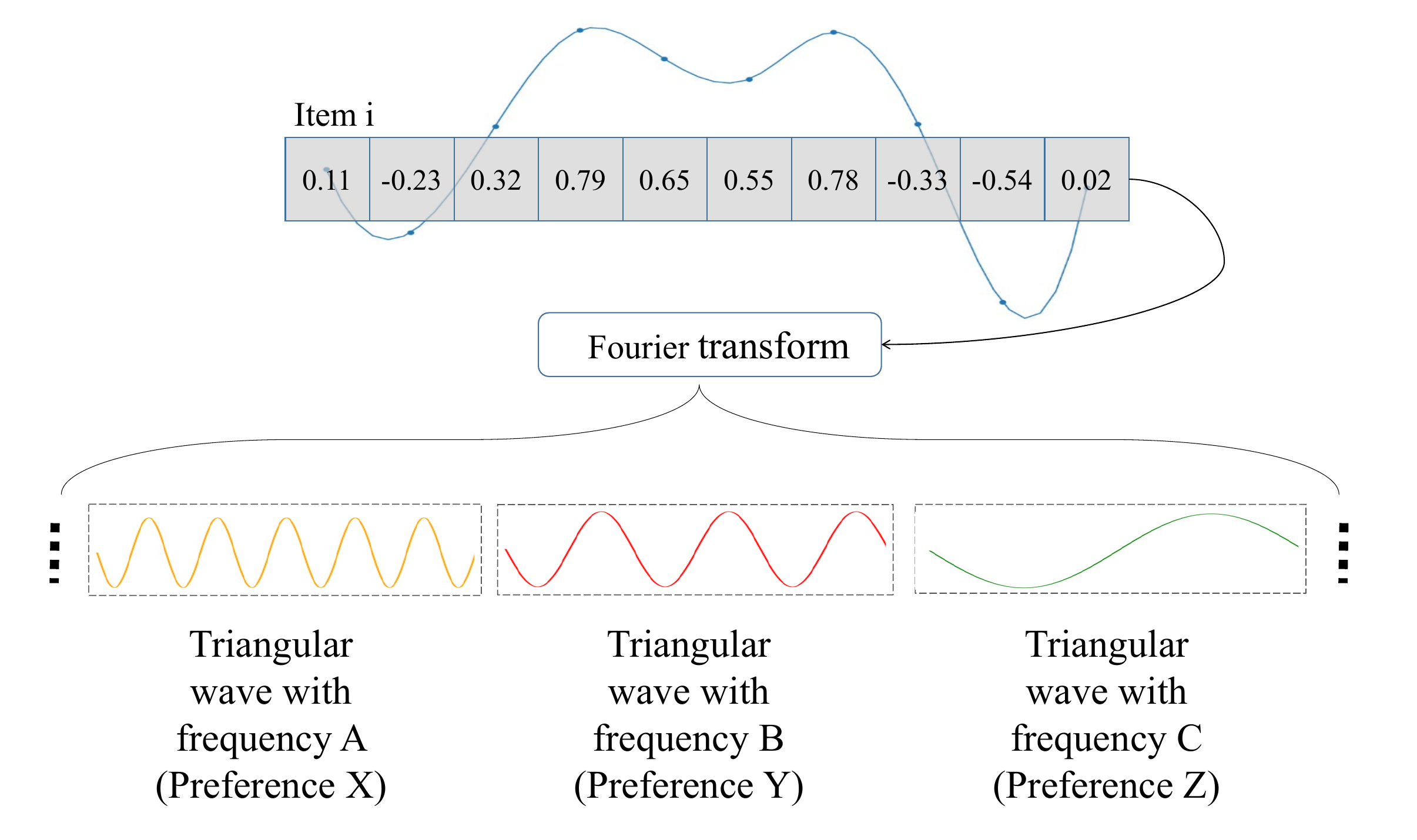}
  \caption{A conceptual illustration of our conjecture: after applying the Fourier transform, we obtain the spectrum of $Item_i$'s embedding at frequencies $\omega_k = \{X, Y, Z\}$. We conjecture that a triangular wave at a specific frequency may loosely correspond to a specific latent preference component $\{A, B, C\}$ -- for instance, $A$ might loosely correspond to a preference for sweet food while $B$ might correspond to an aversion to spicy food. This mapping is illustrative rather than something we claim to have empirically identified.}

  \label{fig:fourier}
 \end{figure}

Discrete Fourier Transform(DFT) is essential in the digital signal processing area \cite{rabiner1975theory,soliman1990continuous} and is a crucial component in our approach. In this paper, we only consider the 1D DFT. Given a sequence of numbers $X =\{x_0, x_1,\cdots, x_{N-1}\}$, the 1D DFT converts the sequence into the frequency domain by the following formulation:
\begin{equation}
    E_{k}=\sum_{n=0}^{N-1} x_{n} e^{-\frac{2 \pi i}{N} n k} 
\end{equation}

where $i$ is the imaginary unit and k$(0 \leq k \leq N-1)$is the index. For each $k$, DFT generates a new representation $E_k$ as a sum of all the original input tokens $x_n$ with so-called “twiddle factors”. In this way, $E_k$ represents the spectrum of the sequence $X$ at the frequency $\omega_k$ = $2\pi$$k$/$N$. Note that DFT is a one-to-one transformation. Given the DFT $E_k$ , we can recover the original sequence $X$ by the inverse DFT:

\begin{equation}
x_{n}=\frac{1}{N} \sum_{k=0}^{N-1} E_{k} e^{\frac{2 \pi i}{N} n k}
\end{equation}
\begin{figure*}[ht]
    \centering
    \includegraphics[width=0.95\textwidth]{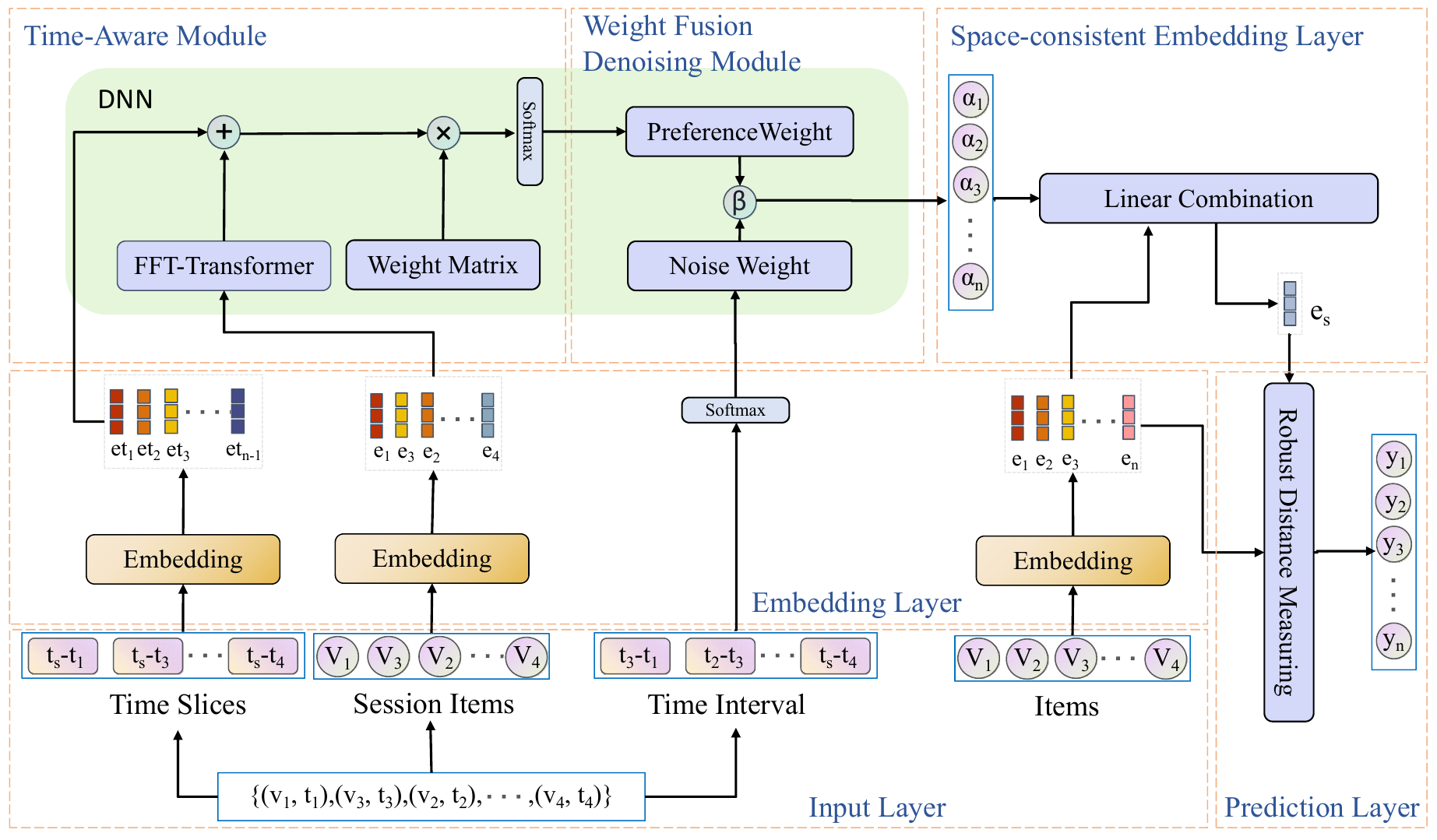}
    \caption{The framework of the DTAMLP}
    \label{fig:model}
\end{figure*}
Our working hypothesis, illustrated in Figure \ref{fig:fourier}, is to treat an item's embedding as if it were a discrete signal. Under this view, after applying the Fourier transform, $E_k$ can be loosely interpreted as a matching score between the embedding and a triangular wave of frequency $\omega_k$; we conjecture that different frequencies may loosely align with different latent, entangled psychological preference components, and that the magnitude of $E_k$ reflects how strongly each such component is expressed. We do not claim to have empirically verified this component-level correspondence; rather, we use it as an intuition for why applying a \emph{learnable} weight matrix on top of the DFT output -- rather than, say, a learnable weight directly in the time domain -- gives the model a natural way to adaptively up- or down-weight different latent preference components across recommendation scenarios. Section~\ref{4.3} operationalizes this intuition as the FFT-Transformer module, and the ablation study in Section~\ref{5-experiments} provides indirect, outcome-level evidence (i.e., that removing this module hurts accuracy) rather than a direct verification of the underlying mechanism.

\section{METHODOLOGY}
\label{4-apporach}

In this section, we describe how DTAMLP integrates the weight fusion mechanism (Section~\ref{3.2.1}) and the FFT-based filtering conjecture (Section~\ref{3-Analysis}) into a single, self-contained system, on top of a representation-consistent embedding design adapted from \cite{hou2022core}. The overall framework of DTAMLP is shown in Figure \ref{fig:model}. The session embedding is obtained by linearly combining all item embeddings within a session, and the weight for each item embedding in the linear combination is generated by a deep neural filtering module, which is the DNN in Figure \ref{fig:model}. This module includes an FFT-Transformer module that applies the frequency-domain filtering intervention discussed in Section~\ref{3-Analysis} to the item embeddings. Moreover, the output of the module combines time embeddings, giving the model time-awareness capabilities. Finally, the weights from the DNN undergo the Weight Fusion module introduced in Section~\ref{3.2.1}, which uses time intervals to reduce the impact of sporadic noise in the model output. The fused weights become the final weights used to linearly combine the various item embeddings within a session, as mentioned earlier. Lastly, the model employs a robust distance-measuring technique, following \cite{hou2022core}, to measure the similarity between the generated session embedding and each item embedding.

We organize the rest of this section as follows: we first illustrate how we obtain embeddings of the input, then introduce the Space-consistent Embedding Module, which we adopt from prior work as the basis of our system. Additionally, we discuss how our DNN generates the weights for linear combinations, i.e., how the two mechanisms from Section~\ref{3-Analysis} are operationalized. At last, a brief explanation of how Robust Distance Measuring works will be presented. 

\subsection{Embedding-layer}\label{4.1}
Each item is embedded into a unified embedding space using an embedding matrix $\mathbf{M}_I \in \mathbb{R}^{|\mathcal{I}| \times d}$ to manage the embedding representations of all items, which maps the one-hot encoding of an item to a d-dimensional vector space. We apply a lookup operation from $\mathbf{M}_I$ to transform an n-length item sequence into an input embedding matrix $\mathbf{E} \in \mathbb{R}^{n \times d}$. Additionally, we define a learnable position encoding matrix $\mathbf{P} \in \mathbb{R}^{n \times d}$ to integrate positional information into the item embeddings. Recent studies have shown that using dropout and layer normalization operations can make training more stable. Ultimately, we generate the embedding representation of each item within a session, $\mathbf{E_i} \in \mathbb{R}^{n \times d}$, by:
\begin{equation}
    \mathbf{E_i} =  \mathbf{\textit{Dropout}(\textit{LayerNorm}(E + P))}
\end{equation}

We obtain all the item embedding through the aforementioned operation. However, the user may interact with different items in different sessions. So the input of our model derives from the given dataset which we divide as Time Slices, Session Items, and Time Intervals. Noted that in terms of Time Slice, We will give a detailed introduction to the embedding of timestamps in the Time-Aware Module Section.

\subsection{Space-consistent Embedding Module} \label{4.2}
\begin{figure}[h]
    \centering
 \includegraphics[width=0.4\textwidth]{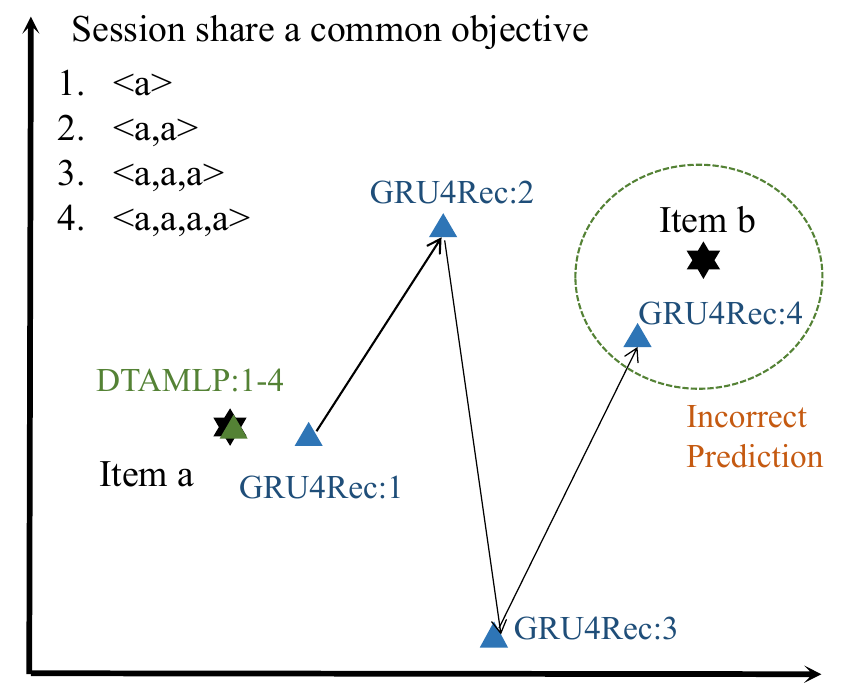}
  \caption{Different representation spaces between session embedding and item embeddings can lead to incorrect predictions when measuring distances between embeddings for recommendation purposes (illustration adapted from the observation in \cite{hou2022core}).}
  \label{fig:rce}
 \end{figure}

 In the context of SBR, session embeddings should reflect users' short-term preferences and be similar to the embeddings of preferred items. However, previous research \cite{hou2022core} raised an interesting discovery that if item embeddings are encoded by non-linear neural networks (such as RNN-based), the ultimate session embedding and item embedding may be in different representation spaces, as illustrated in Figure~\ref{fig:rce}. This design choice is not a contribution of this paper; we adopt it directly from \cite{hou2022core} because it is important for our system-level implementation: subsequent evaluation methods based on vector similarity comparison may lose their theoretical basis, contributing to the model's incorrect recommendation, if this inconsistency is not addressed. We address this issue with the following modules.

\paragraph{Representation-Consistent Encoding}

To ensure that the session embedding and item embeddings are represented in the same space, we design the session embedding as a weighted summary of the item embeddings within a session. 

\begin{equation}
    \mathbf{\alpha} =  DNN([\mathbf{h_{s,1};h_{s,2};...;h_{s,n}])}  
\end{equation}
\begin{equation}
    \mathbf{h_s} = \sum_{\boldsymbol{i}=1}^n {\alpha}_i\mathbf{h_{s,i}}
\end{equation}

Specifically, we employ a Deep Neural Filtering Blocks structure with neural filtering capabilities to learn weights for each item embedding within a session, which can fully mine implicit features in data. The details about this structure are shown in the following section.

\subsection{Deep Neural Filtering Blocks} \label{4.3}
\paragraph{FFT-Transformer}
In this module, we employ an improved DFT operation to compute the weight for each item in a session, as opposed to the multi-head mechanism utilized in the original Transformer model \cite{vaswani2017transformer}. Let $\mathbf{F^l} \in \mathbb{R}^{n \times d}$ as the $i_{th}$  layer of the input item representation matrix. Firstly, We perform FFT along the item dimension to transform $\mathbf{F^l}$ into the frequency domain:
\begin{equation}
    \widetilde{\mathrm{F}}^l =\mathcal{F}^{-1}\left( \mathrm{W} \odot 
 \mathcal{F}\left(\mathrm{F}^l\right) \right)
\end{equation}

Here, $\mathcal{F\left(\right)}$ denotes the one-dimensional FFT, and transform $F_l$ into spectrum. We can then modulate the spectrum by multiplying a learnable filter $W$, which can be optimized by SGD to adaptively represent an arbitrary filter in the frequency domain. Finally, we adopt the inverse FFT to transform the modulated spectrum back to the time domain and update the sequence representations. $\mathcal{F}^{-1}\left(\right)$ denotes the inverse 1D FFT, which converts the complex tensor into a real number tensor. With the operations of FFT and inverse FFT, the preference noise can be effectively reduced. Following SASRec 
\cite{kang2018sasrec}, we also incorporate the skip connection \cite{he2016deep}, layer normalization \cite{ba2016layer}, and dropout 
 \cite{srivastava2014dropout} operations to alleviate the gradient vanishing and unstable training problems as:
\begin{equation}
    \widetilde{\mathbf{F}}^l=\text { LayerNorm }\left(\mathrm{F}^l+\operatorname{Dropout}\left(\widetilde{\mathbf{F}}^l\right)\right)
\end{equation}
In the point-wise feed-forward network, we incorporate MLP and ReLU activation functions to further capture the non-linearity characteristics. The computation is defined as: 
\begin{equation}
\operatorname{FFN}\left(\widetilde{\mathbf{F}}^l\right)=\left(\operatorname{ReLU}\left(\widetilde{\mathbf{F}}^l \mathbf{W}_1+\mathbf{b}_1\right)\right) \mathbf{W}_2+\mathbf{b}_2
\end{equation}
where $W_1$, $b_1$, $W_2$, $b_2$ are both trainable parameters. Then, we perform skip connection and layer normalization.

\paragraph{Time-Aware Module}
Through FFT-Transformer, item embeddings acquire implicit relationships and features and we simultaneously filter out potential preference noise. To further enrich the item embeddings with additional dimensions of information, we incorporate time information into the embeddings. Specifically, we compute the time embedding $et_{i}$ using the following formula:
\begin{equation}
    e t_i=e m b\left(\min \left(\left\lfloor\frac{t_c-t_i}{600}\right\rfloor, 144\right)\right)
\end{equation}
where $t_i$ refers to the click timestamp of the session item, and $t_c$ denotes the prediction timestamp. Dividing the time interval $t_c$ - $t_i$ by 600 indicates that the time slice is 10 minutes, which is determined experimentally. $\lfloor \cdot \rfloor$ is the rounding operation to obtain an integer from a real number. This discretization process significantly reduces the number of time embedding vectors.  $e m b$($x$) is a linear embedding function used to transform the time feature into a low-dimensional latent vector; 144 is the threshold of the number of time slices, namely one day (144*600s)
We integrated time information into item embeddings using the following formula:
\begin{equation}
    e = FFT{-}Transformer(\mathrm{F}^l)
\end{equation}
\begin{equation}
    \alpha_{temp} = H^T(e||et_{i})
\end{equation}

Here, $e$ is the item embedding, which is the output of the FFT-Transformer Module. Item embedding and time embedding are concatenated, and a weight matrix reduces the resulting 2 $\times$ hidden size vector to a scalar value $ \alpha_{temp}$. 

\paragraph{Weight Fusion Module}
The Time-aware module above incorporates time information into the weight output of the DNN, but does not by itself distinguish sporadic noise from genuine interest, since it still learns an attention weight over every click. We therefore integrate the weight fusion mechanism validated in Section \ref{3.2.1} into DTAMLP. Recall that in Section \ref{3.2.1}, we compute an \textit{interval} feature as the time difference between consecutive item clicks and cap it with a hyperparameter $\eta$. We use the following formula to compute the final weights $\alpha$ of the items:
\begin{equation}
    \alpha_{noise} =\operatorname{softmax}\left(\min \left(interval, \eta \right)\right)
\end{equation}
\begin{equation}
    \alpha=\left(1 -\beta \right) \times  \alpha_{temp} +  \beta \times  \alpha_{noise} 
\end{equation}
$ \alpha_{temp}$ refers to the weight generated by the FFT-Transformer module as well as combined with time information, while $\alpha_{noise}$ denotes the weight generated based on the time noise we calculated. $\beta$ is blend coefficient, which is a learnable parameter. A smaller $\alpha_{noise}$ weight indicates a higher likelihood of an item being sporadic noise. After weight fusion, the influence of such an item in generating the session embedding through weighted summation is effectively reduced.

\subsection{Robust Distance Measure} \label{4.4}
Following the setting commonly used in previous works \cite{chen2020simclr,hou2022core}, we utilize cosine distance to facilitate better alignment and uniformity of item embeddings. Besides, we leverage the lemma and definition established in \cite{hou2022core} and approximate the optimization of dot product cross-entropy as optimizing the (N-1)-tuplet loss with a fixed margin of $\tau$. The loss function is referred to as Robust Distance Measuring (RDM), which incorporates the aforementioned modifications.
\begin{equation}
    \ell=-\log \frac{\exp \left(\cos \left(\boldsymbol{h}_s, \boldsymbol{h}_{v^{+}}^{\prime}\right) / \tau\right)}{\sum_{\boldsymbol{i}=1}^m \exp \left(\cos \left(\boldsymbol{h}_s, \boldsymbol{h}_{v_i}^{\prime}\right) / \tau\right)}
\end{equation}
where $\boldsymbol{h}^{\prime}$ denotes the item embeddings with dropout.

\section{EXPERIMENT}
\label{5-experiments}
\begin{table}[ht]
    \centering
    \caption{Statistics of the two datasets used for the main quantitative comparison in Table~\ref{tab:performance_diginetica} and Table~\ref{tab:performance_retailrocket}. Yoochoose is not included here since it is only used for the qualitative visualization in Section~\ref{5-experiments}.}
    \label{tab:dataset}
    \setlength{\abovecaptionskip}{1pt}
    \setlength{\belowcaptionskip}{1pt}
    \begin{tabular}{lrr}
    \toprule
        \textbf{Statistic} & \textbf{Diginetica} & \textbf{RetailRocket}  \\
    \midrule
        \# Interactions & 789{,}641 & 1{,}061{,}109  \\
        \# Items & 42{,}862 & 61{,}069  \\
        \# Training sessions & 163{,}831 & 274{,}220  \\
        \# Test sessions & 20{,}329 & 33{,}721 \\
        Avg. session length & 4.86 & 4.1 \\
    \bottomrule
    \end{tabular}
\end{table}

\begin{table*}[t]
\vspace{-0.01cm}
	\centering
    \setlength{\abovecaptionskip}{1pt}
	\setlength{\belowcaptionskip}{1pt}
		\caption{Overall performance comparison on \textbf{Diginetica}. Sessions are split into train/validation/test sets in a ratio of 8:1:1 for a fair evaluation. ``MRR@K'' is the Top-k Mean Reciprocal Rank. ``NDCG@K'' is the Top-k Normalized Discounted Cumulative Gain. ``Recall@K'' is the Top-k Recall ratio. Models are grouped by architecture family, top to bottom: All-MLP, Attention-based, GNN-based, RNN-based, CNN-/Gated-based. FMLP-Rec is the all-MLP baseline discussed in Section~\ref{3-Analysis}. Best results are in \textbf{bold}.}
		\label{tab:performance_diginetica}
\setlength{\tabcolsep}{3mm}
\renewcommand\arraystretch{1.1}
\begin{tabular}{lcccccc}
    \toprule
    \textbf{Model} & \textbf{MRR@10} & \textbf{MRR@20} & \textbf{NDCG@10} & \textbf{NDCG@20} & \textbf{Recall@10} & \textbf{Recall@20} \\
    \midrule
    \textbf{DTAMLP} & \textbf{17.92} & \textbf{18.79} & \textbf{23.05} & \textbf{26.30} & \textbf{40.26} & \textbf{53.10} \\
    \midrule
    FMLP-Rec \footnotesize{(All-MLP)} & 16.87 & 17.69 & 20.86 & 24.11 & 37.24 & 50.16 \\
    \midrule
    TiSASRec \footnotesize{(Attn.)} & 16.21 & 17.12 & 21.04 & 24.34 & 36.87 & 49.82 \\
    SASRec \footnotesize{(Attn.)} & 14.29 & 15.06 & 18.44 & 21.59 & 32.39 & 44.85 \\
    Bert4Rec \footnotesize{(Attn.)} & 10.73 & 11.47 & 14.29 & 16.98 & 26.01 & 36.63 \\
    \midrule
    GCARM \footnotesize{(GNN)} & 16.95 & 17.84 & 21.97 & 25.23 & 38.41 & 51.24 \\
    GCSAN \footnotesize{(GNN)} & 15.83 & 16.69 & 20.51 & 23.65 & 35.82 & 48.31 \\
    SR-GNN \footnotesize{(GNN)} & 14.95 & 15.82 & 19.61 & 22.78 & 34.87 & 47.47 \\
    \midrule
    GRU4Rec \footnotesize{(RNN)} & 14.71 & 15.59 & 19.17 & 22.39 & 33.81 & 46.58 \\
    NARM \footnotesize{(RNN)} & 14.86 & 15.73 & 19.52 & 22.71 & 34.82 & 47.45 \\
    \midrule
    Caser \footnotesize{(CNN)} & 11.92 & 12.71 & 15.91 & 18.82 & 29.04 & 40.58 \\
    HGN \footnotesize{(Gated)} & 11.45 & 12.21 & 15.41 & 18.22 & 28.42 & 39.49 \\
    \bottomrule
\end{tabular}
\vspace{-0.2cm}
\end{table*}

\begin{table*}[t]
\vspace{-0.01cm}
	\centering
    \setlength{\abovecaptionskip}{1pt}
	\setlength{\belowcaptionskip}{1pt}
		\caption{Overall performance comparison on \textbf{RetailRocket}, using the same experimental protocol, metrics, and architecture-family grouping as Table~\ref{tab:performance_diginetica}.}
		\label{tab:performance_retailrocket}
\setlength{\tabcolsep}{3mm}
\renewcommand\arraystretch{1.1}
\begin{tabular}{lcccccc}
    \toprule
    \textbf{Model} & \textbf{MRR@10} & \textbf{MRR@20} & \textbf{NDCG@10} & \textbf{NDCG@20} & \textbf{Recall@10} & \textbf{Recall@20} \\
    \midrule
    \textbf{DTAMLP} & \textbf{43.29} & \textbf{43.64} & \textbf{46.92} & \textbf{48.29} & \textbf{58.74} & \textbf{64.23} \\
    \midrule
    FMLP-Rec \footnotesize{(All-MLP)} & 41.44 & 41.84 & 44.75 & 45.87 & 57.63 & 63.31 \\
    \midrule
    TiSASRec \footnotesize{(Attn.)} & 41.32 & 41.76 & 45.11 & 46.71 & 57.22 & 63.49 \\
    SASRec \footnotesize{(Attn.)} & 36.40 & 36.82 & 39.81 & 41.31 & 50.68 & 56.62 \\
    Bert4Rec \footnotesize{(Attn.)} & 26.05 & 26.38 & 28.93 & 30.14 & 38.14 & 42.92 \\
    \midrule
    GCARM \footnotesize{(GNN)} & 41.03 & 41.41 & 45.07 & 46.44 & 57.82 & 63.25 \\
    GCSAN \footnotesize{(GNN)} & 38.30 & 38.65 & 41.97 & 43.19 & 53.31 & 58.37 \\
    SR-GNN \footnotesize{(GNN)} & 38.57 & 38.95 & 42.34 & 43.74 & 54.29 & 59.79 \\
    \midrule
    GRU4Rec \footnotesize{(RNN)} & 38.42 & 38.84 & 42.03 & 43.41 & 53.50 & 58.94 \\
    NARM \footnotesize{(RNN)} & 38.61 & 39.12 & 42.26 & 43.67 & 53.81 & 59.41 \\
    \midrule
    Caser \footnotesize{(CNN)} & 25.58 & 25.95 & 28.89 & 30.34 & 39.84 & 45.14 \\
    HGN \footnotesize{(Gated)} & 25.23 & 25.62 & 28.93 & 30.28 & 40.66 & 45.99 \\
    \bottomrule
\end{tabular}
\vspace{-0.2cm}
\end{table*}
\subsection{Experimental Setup}

\subsubsection{Dataset}
To evaluate the efficacy of our proposed method, we use two publicly available datasets, Diginetica and RetailRocket, for the main quantitative comparison in Table~\ref{tab:performance_diginetica} and Table~\ref{tab:performance_retailrocket}, and the ablation study. We additionally use Yoochoose for a qualitative t-SNE visualization in Section~\ref{5-experiments} only, since we did not carry out the full quantitative evaluation on it. These datasets have varying domains, sparsity levels, and sizes, and are open access.
\begin{itemize}
    \item {\textbf{Diginetica:} The Diginetica dataset is obtained from the CIKM Cup 2016 and the experiments only need the transaction data part}
    \item {\textbf{Retailrocket:} RetailRocket is collected from a personalized e-commerce website. It contains six months of user browsing activities. Only the behavior data part (events.csv) is used in the experiments}
    \item {\textbf{Yoochoose:} The Yoochoose dataset is obtained from the RecSys Challenge 2015, which captures user clicks on an e-commerce site in a sequence. As noted above, we only use it for the qualitative visualization in Section~\ref{5-experiments}.}
\end{itemize}  

In order to ensure a fair comparison with existing methods, sessions with a length of one and items with less than five occurrences were removed. Following the approach in \cite{hou2022core}, the click sequences in Diginetica and RetailRocket were divided into sessions using a certain ratio of 8:1:1, respectively. Since some of the sessions in the RetailRocket dataset are too long to be used, only the last 50 clicks in each session were retained after processing. Each session was then split into a training sequence and a label. Specifically, for a given input session $s=\{v_1, v_2,\cdots, v_n\}$, the following training sequence was generated: [$\left(v_1, y\left(v_2\right)\right)$, $\left(v_1, v_2, y\left(v_3\right)\right)$, $\left(v_1, v_2, v_3, y\left(v_4\right)\right)$, $...$, $\left(v_1, v_2, v_3, ..., v_{n-1}\right)$, $y\left(v_n\right)$ ]. Here,  $y\left(v_i\right)$ represents the item to be clicked next in the session sequence. Due to the largeness of the Yoochoose dataset, we selected the most recent 1/64 sequence fragments as the training data. Finally, items that only appeared in the test set were filtered out. Table \ref{tab:dataset} provides statistical information on the datasets used in the experiment. 

\subsection{Experiment Details}
The dimension of the latent vectors is fixed to 100, and each session is truncated within a maximum length of 50. We optimized all the compared methods using Adam optimizer \cite{kingma2015adam} with a learning rate of 0.001. We use a batch size of 2048 for all methods. Other hyper-parameters of baselines are carefully tuned following the suggestions from the original papers and we report each performance under its optimal settings. 

\subsubsection{Baseline Models} 
We compared the performance with following methods: 
(1)RNN-based methods: GRU4Rec \cite{hidasi2016gru4rec}, NARM \cite{li2017narm} (2)Graph-based methods: SR-GNN \cite{wu2019srgnn}, GCSAN \cite{xu2019gcsan}, GCARM \cite{pan2021graph} (3)CNN-based methods: Caser\cite{tang2018personalized} (4)Gated-based: HGN \cite{ma2019hierarchical} (5)Attention-based methods: TiSASRec \cite{li2020time}, Bert4Rec
\cite{sun2019bert4rec}, SASRec \cite{kang2018sasrec} (6)All-MLP method: FMLP-Rec \cite{zhou2022fmlp}, which is the model whose frequency-domain filtering behavior directly motivated our preference-noise conjecture in Section~\ref{3-Analysis}.
Most baselines are implemented based on the popular open-source recommendation library, and codes of RecBole and GCARM are provided by the author. In this experiment, we employed widely used evaluation metrics such as top-k Recall ratio (Recall@k), Top-k Mean Reciprocal Rank (MRR@k), and Top-k Normalized DCG (NDCG@k).

\subsection{Experimental Results}

\subsubsection{Overall comparison} 
Our experimental results are presented in Table \ref{tab:performance_diginetica} (Diginetica) and Table \ref{tab:performance_retailrocket} (RetailRocket). From the tables, we can draw several conclusions. Firstly, it is worth mentioning that GNN-based baselines such as GCARM and SR-GNN perform well, which shows the general competitiveness of graph-based models on these two datasets. The all-MLP baseline FMLP-Rec is also competitive, in particular on RetailRocket, which is consistent with our discussion in Section~\ref{3-Analysis} that frequency-domain filtering already captures a useful, if unexplained, signal.

Comparing DTAMLP against all baseline models, we observe consistent improvements on both datasets. On Diginetica, the strongest baseline across all six metrics is GCARM, and DTAMLP improves over it by 5.72\%, 5.32\%, 4.92\%, 4.24\%, 4.82\%, and 3.63\% (relative improvement) on MRR@\{10,20\}, NDCG@\{10,20\}, and Recall@\{10,20\}, respectively. On RetailRocket, the strongest baseline differs by metric (FMLP-Rec on MRR, TiSASRec on NDCG and Recall@20, GCARM on Recall@10); relative to the best baseline for each metric, DTAMLP improves MRR@\{10,20\} by 4.46\% and 4.30\%, NDCG@\{10,20\} by 4.01\% and 3.38\%, and Recall@\{10,20\} by 1.59\% and 1.17\%. We note that the Recall gains on RetailRocket are noticeably smaller than the gains on the ranking-sensitive metrics (MRR, NDCG), suggesting that DTAMLP's improvement mainly comes from ranking relevant items higher rather than retrieving a substantially different set of relevant items. We further present parameter tuning results in Section~\ref{5-experiments} and analyze our model's efficiency below.
\begin{figure}[ht]
		\setlength{\abovecaptionskip}{1pt}
		\setlength{\belowcaptionskip}{1pt}
		\centering
  		\subfigure[GCARM]{
			\includegraphics[width=0.3\columnwidth]{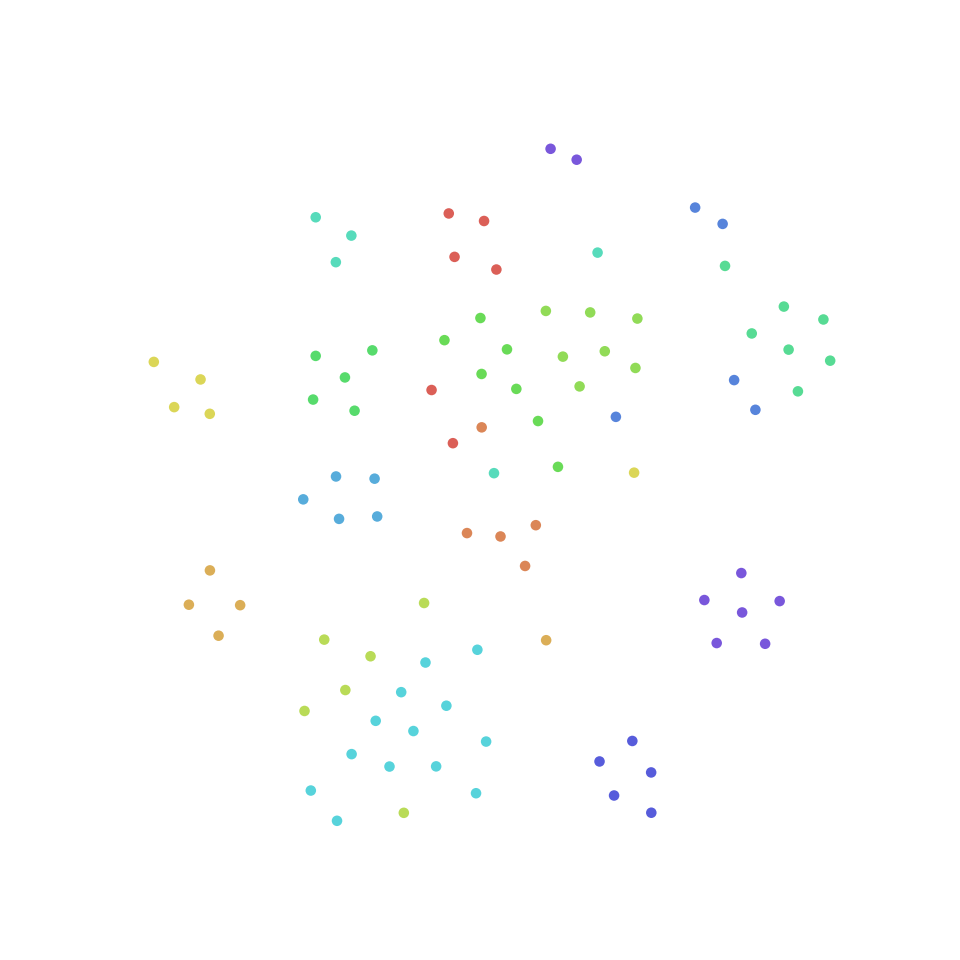}
			\label{yoochoose1:GCARM}
		}%
		\subfigure[SR-GNN]{
			\includegraphics[width=0.3\columnwidth]{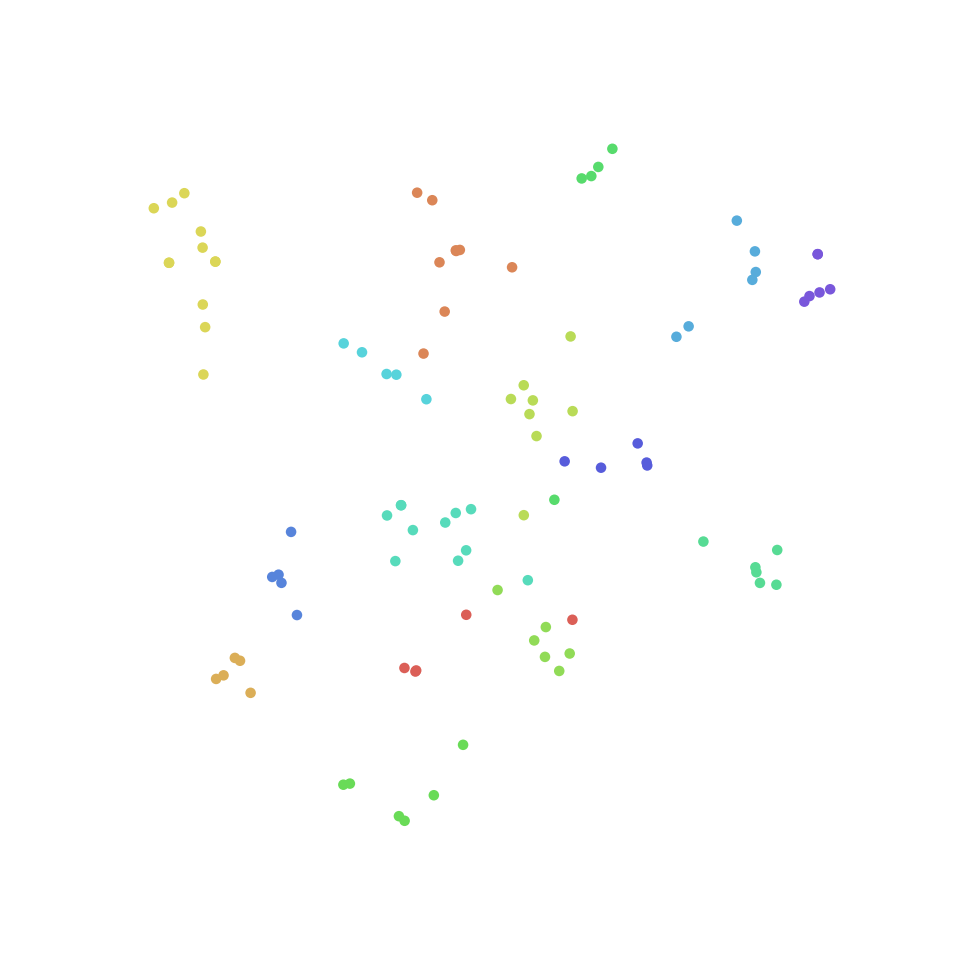}
			\label{yoochoose1:SRGNN}
		}%
		\subfigure[DTAMLP]{
			\includegraphics[width=0.3\columnwidth]{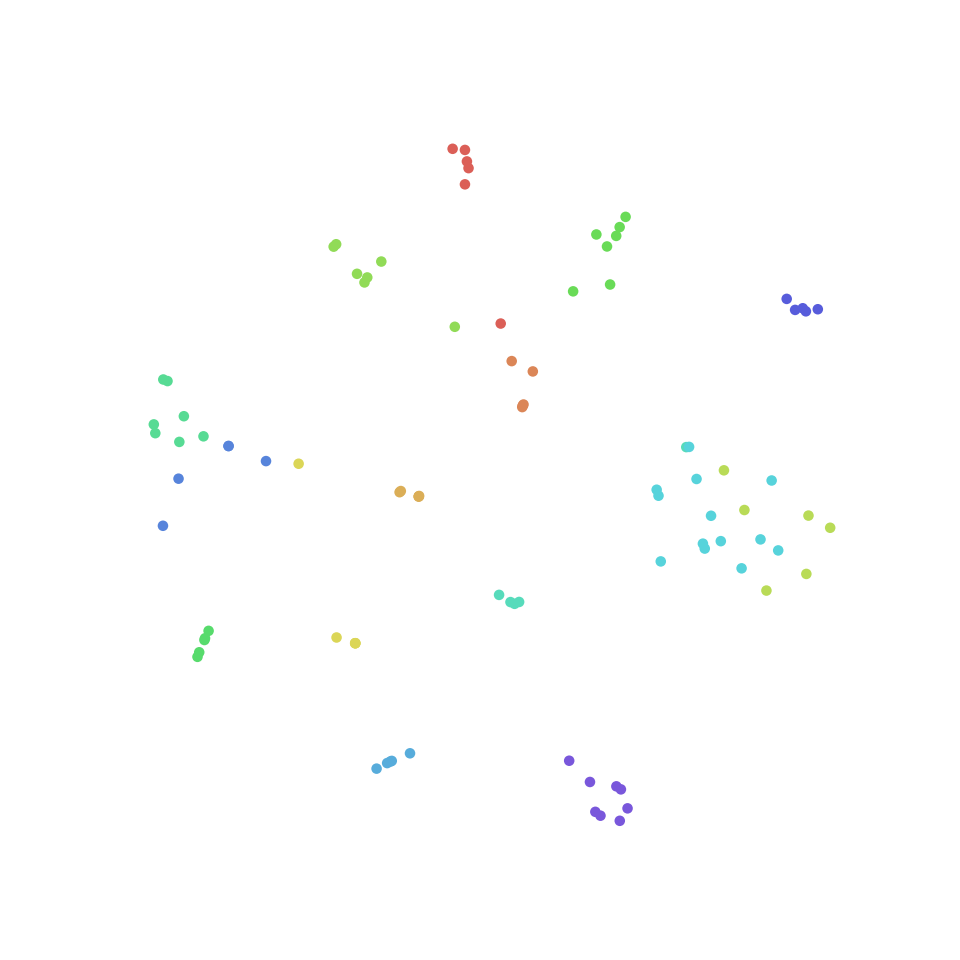}
			\label{yoochoose1:DTAMLP}
		}
  \caption{Visualization of learned session embeddings on Yoochoose datasets}
		\label{fig:embedding}
\vspace{-0.4cm}
\end{figure}

\subsubsection{Ablation study}
\begin{figure}[ht]

		\setlength{\abovecaptionskip}{1pt}
		\setlength{\belowcaptionskip}{1pt}
		\centering
  		\subfigure[Retailrocket]{
			\includegraphics[width=0.23\textwidth]{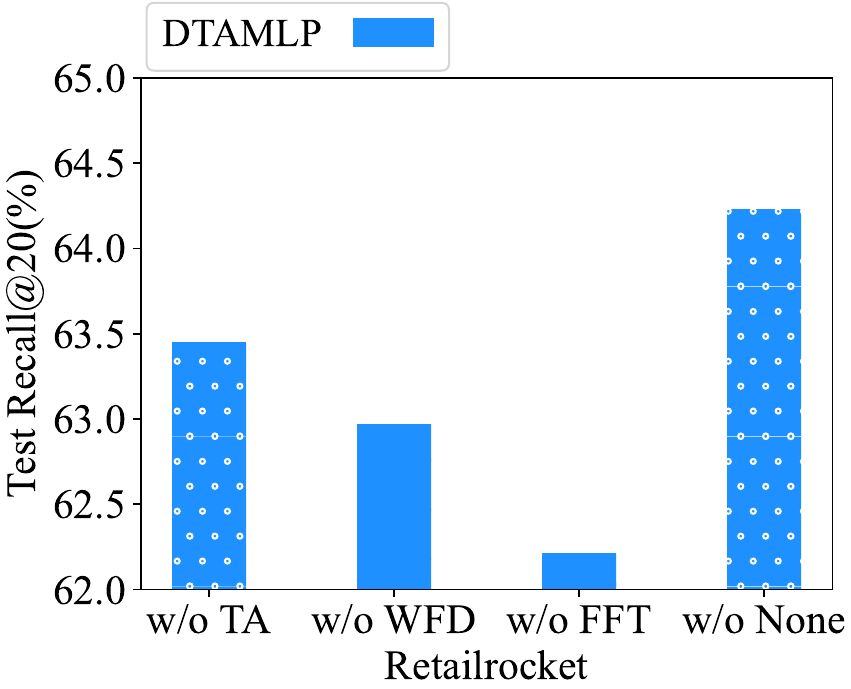} 
			\label{retailrocket-abla}
		}
    \hspace{-0.4cm}
		\subfigure[Diginetica]{
			\includegraphics[width=0.23\textwidth]{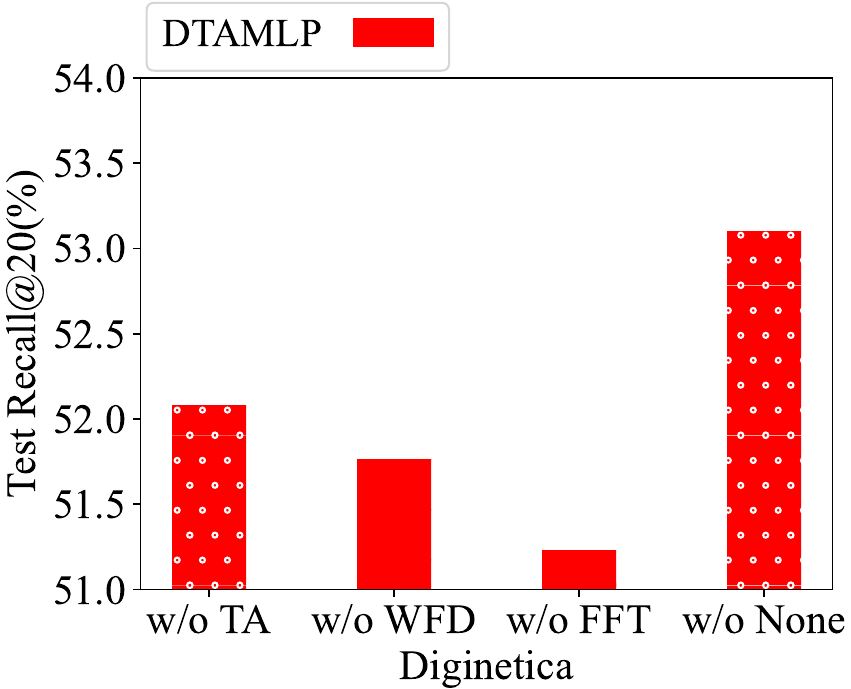}
			\label{diginetica-abla}
		}
  \caption{Ablation study of DTAMLP on Diginetica and Retailrocket. See Section~\ref{5-experiments} for the definition of each variant.}
\label{fig:ablation1}
\vspace{-0.4cm}
\end{figure}

We conduct ablation experiments on Diginetica (using MRR) and Retailrocket (using Recall) by removing the three key components one at a time. Results are shown in Figure \ref{fig:ablation1}, where ``w/o TA'' denotes no time-aware ability, ``w/o WFD'' denotes no usage of the Weight Fusion module (weights are generated solely from the FFT-Transformer output), ``w/o FFT'' denotes replacing the FFT-Transformer with the original Transformer, and ``w/o None'' is the full DTAMLP with none of those parts removed. Comparing ``w/o WFD'' (and ``w/o TA'') against the full model isolates the contribution of Insight 1 (sporadic-noise weight fusion), while ``w/o FFT'' isolates Insight 2 (frequency-domain filtering). Removing any of the three components impacts accuracy, and the two components tied to our central insights both cause a clear drop relative to the full model, with the FFT-Transformer module having a relatively larger impact overall. Since removing either component independently hurts performance and removing both (implicitly, by comparing against the individual-baseline results in Table~\ref{tab:performance_diginetica} and Table~\ref{tab:performance_retailrocket}) hurts it further, we take this as ablation-level evidence that the two mechanisms contribute complementary, non-redundant improvements, consistent with the way they were motivated separately in Section~\ref{3-Analysis}.

\subsubsection{Parameter tuning} 
\begin{figure}[t]
		\setlength{\abovecaptionskip}{1pt}
		\setlength{\belowcaptionskip}{1pt}
		\centering
  \vspace{0.1cm}
  		\subfigure[Item Dropout]{
			\includegraphics[width=0.22\textwidth]{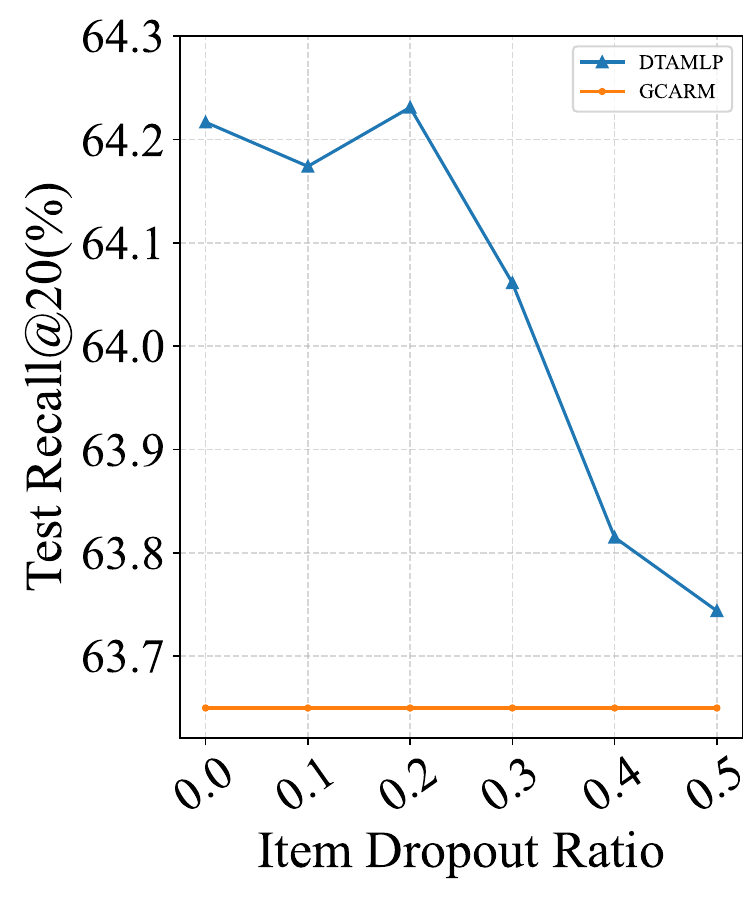} 
			\label{item-dropout}
		}
    \hfill
		\subfigure[Margin]{
			\includegraphics[width=0.22\textwidth]{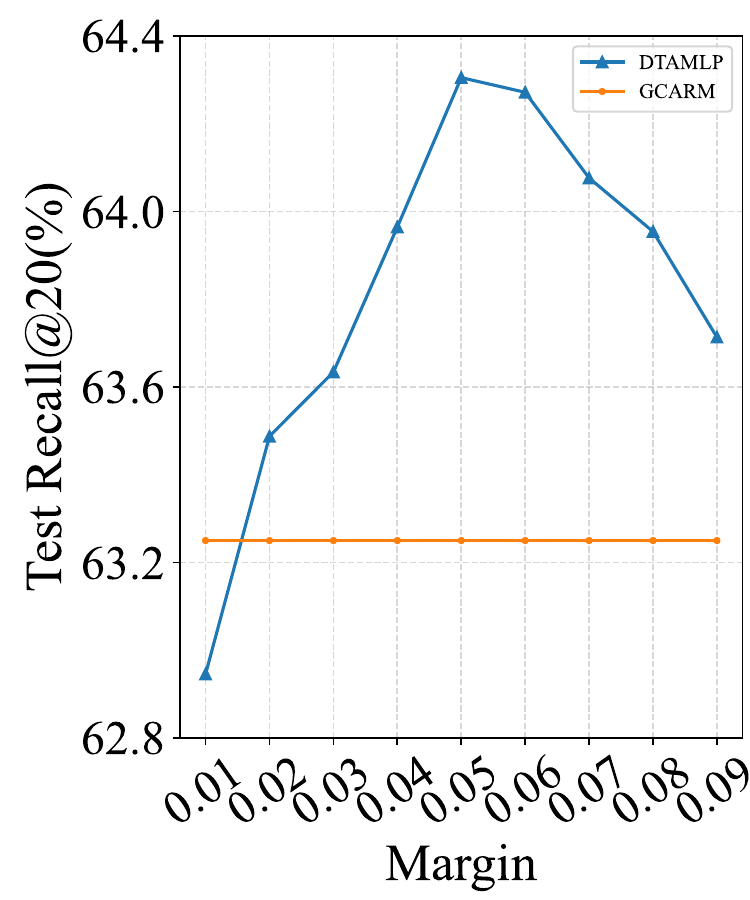}
			\label{Margin}
		}
    \\
  \vspace{0.05cm}
  		\subfigure[Time Interval]{
			\includegraphics[width=0.22\textwidth]{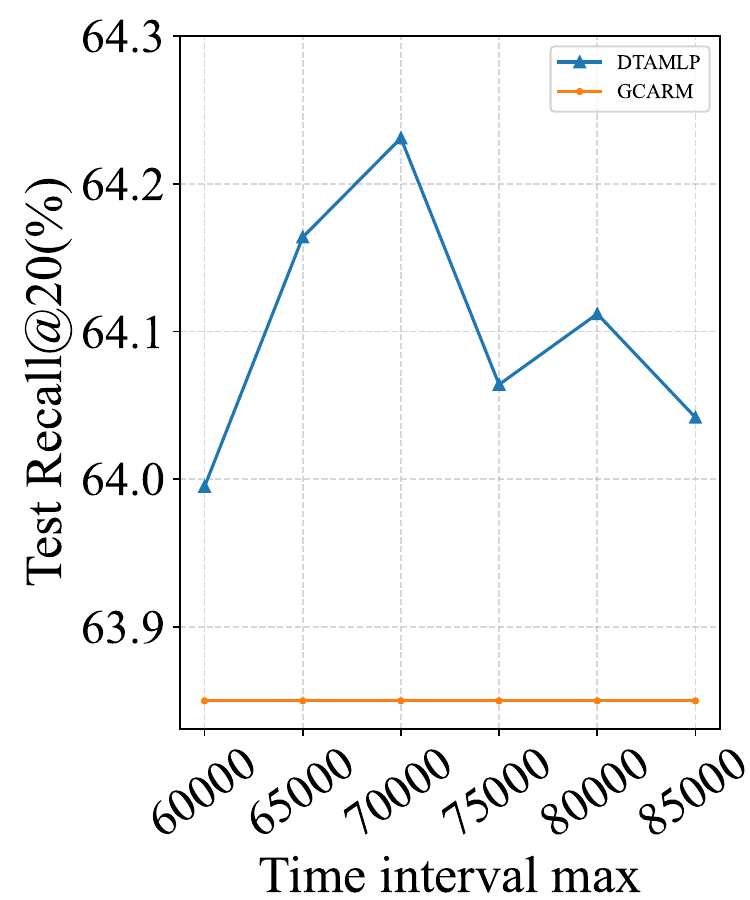} 
			\label{time-interval}
		}
    \hfill
		\subfigure[Embedding Size]{
			\includegraphics[width=0.22\textwidth]{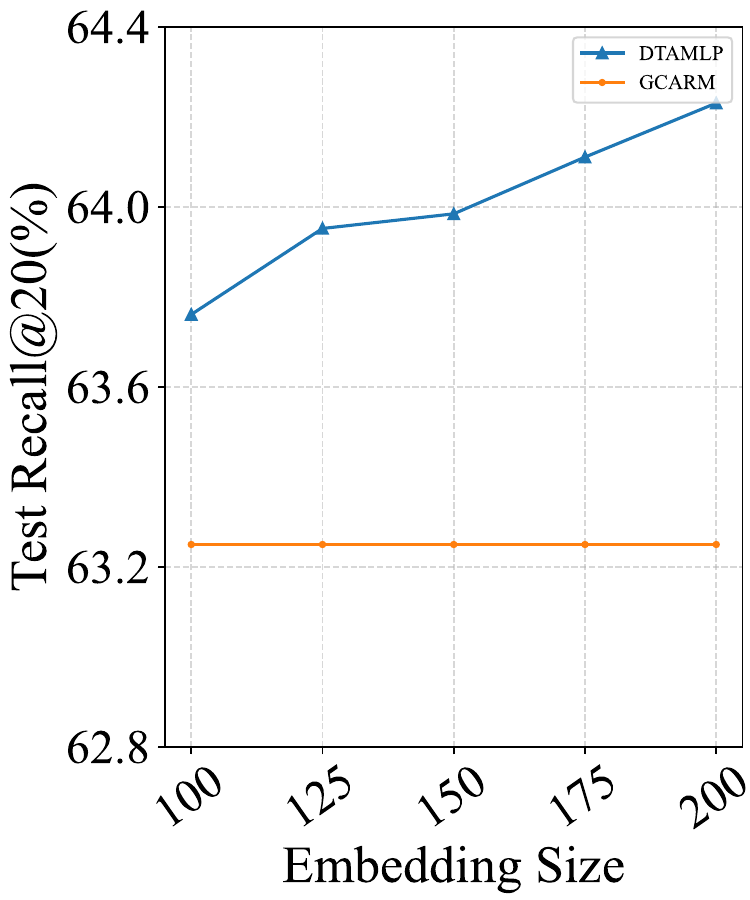} 
			\label{Embedding Size}
		}
  \caption{Parameter tuning of DTAMLP on Retailrocket datasets.}
   \label{fig:ablation}
\vspace{-0.1cm}
\end{figure}

In the final part of our experiment, we focused on validating several crucial parameters that impact the performance and sensitivity of DTAMLP. These parameters include $interval_{max}$, dropout ratio, $slice_{max}$, margin, and embedding size. To be more specific, we varied the value of $interval_{max}$ in \{70000, 75000, 80000, 65000, 60000, 85000, 90000\} and present the results in Figure \ref{fig:ablation}. Our model consistently outperforms the state-of-the-art and achieves the best experimental results on the Retailrocket dataset when $interval_{max}$ is equal to 70000.

Furthermore, we varied the value of the margin in \{0.01, 0.02, 0.03, 0.04, 0.05, 0.06, 0.07, 0.08, 0.09\}. The experiment revealed that a margin value of 0.05 leads to the best performance of DTAMLP on Retailrocket. Additionally, a small margin value has a significant impact on the performance of the model and even causes its performance to fall below that of the GCARM model. Next, we varied the dropout ratio in the range of 0 to 0.5 and found that a dropout ratio of 0.2 yields the greatest improvement in model performance. Finally, we tested the effect of embedding size on the model's performance. We find that when the embedding size gets larger, the performance gets better accordingly. But it will cost more GPU memory.

\subsubsection{Visualization of session embeddings} 
As a qualitative, illustrative complement to the quantitative results above, we present a visualization of the learned session embeddings using the t-SNE \cite{van2008visualizing} algorithm in Figure \ref{fig:embedding}, on a subset of the Yoochoose dataset (this dataset is used here only for this qualitative visualization and is not part of the main quantitative comparison in Table~\ref{tab:performance_diginetica} and Table~\ref{tab:performance_retailrocket}). Specifically, we group sessions with the same next item and represent them with the same color to facilitate comparison. To establish ground truth, we randomly select 15 items and extract all corresponding sessions from our test set. Notably, the session embeddings learned by DTAMLP exhibit visibly better class separation than those derived from SR-GNN and GCARM, which is consistent with -- though not a formal proof of -- our claim that DTAMLP produces a more discriminative session representation.

\subsection{Efficiency Analysis}
In this section, we compare three representative recommendation system models in terms of time complexity: SR-GNN, TiSASRec, and Caser. We selected these three models for comparison because GCARM is a GNN-based model, TiSASRec is a time-aware attention-based model, and Caser is a CNN-based model, each representing different neural network architectures. We use a sequence of length $d$ to measure the time complexity of different models, to avoid differences in algorithms between models (such as dropout and normalization layers).

Firstly, SR-GNN uses GNN, which requires $O(d^2)$ time complexity for information aggregation and forgetting in one layer. If there are n layers of GNN, it will be $O(nd^2)$. TiSASRec is a self-attention mechanism model, with a time complexity of $O(d^2h)$ for both similarity calculation and weighted sum, where $h$ is the hidden layer dimension. Caser is CNN-based, and for the input size $d \times h$, the complexity of one operation for a convolution kernel of size $k \times h$ is $O(kh)$. It is done d times in total, resulting in a complexity of $O(dkh)$. For our DTAMLP, the FFT-Transformer module (Insight 2) replaces multi-head attention with a discrete Fourier transform, giving a theoretical time complexity of $O(d\log d)$. Among all baselines, Caser is the only model with a theoretically lower complexity than DTAMLP; however, in our experiments we could not use a large batch size for Caser due to its limited parallelizability (e.g., it frequently produced NaN values at higher batch sizes). In practice, Caser took nearly 6 hours to train on a single Nvidia V100 16G GPU, whereas DTAMLP took less than 10 minutes, and Caser's accuracy is also far below that of DTAMLP (Table~\ref{tab:performance_diginetica} and Table~\ref{tab:performance_retailrocket}). We did not measure training time for GCARM, the strongest GNN-based baseline in our accuracy comparison, so we do not claim an efficiency advantage over it; the comparison above is intended to show that replacing attention/GNN aggregation with an FFT-based module is a reasonable, low-complexity design choice rather than to claim overall superiority in efficiency.

\section{CONCLUSION}
\label{6-conclusion}
This paper reports two empirical observations on session-based recommendation that we made in 2023, together with a system, DTAMLP, that integrates them. The central and most directly verifiable finding is that a small, plug-and-play weight fusion intervention -- fusing an existing model's attention weight with a simple, threshold-capped function of the click-time interval -- can be inserted into off-the-shelf time-aware and GNN-based models such as TiSASRec and SR-GNN with almost no architectural change, yet yields a consistent accuracy improvement (Table~\ref{table:ti}). We refer to the underlying phenomenon as sporadic noise. The second observation is more speculative: motivated by the under-explained effectiveness of frequency-domain filtering in FMLP-Rec, we offer a conjecture -- which we call preference noise -- for why separating a user's entangled psychological preferences may be easier in the frequency domain than in the time domain; we present this as an interpretive hypothesis rather than a proven mechanism. Combining both observations with a representation-consistent embedding design adapted from prior work, we built DTAMLP and validated it on Diginetica and RetailRocket, with ablation results indicating that the two mechanisms contribute complementary, non-redundant improvements.

We are releasing this work primarily to document these two observations and the reasoning behind them, rather than to claim a current state-of-the-art system: the DTAMLP results reflect the state of the field circa 2023, and we have not updated the experiments or baselines since. We believe the plug-and-play weight fusion result (Insight 1) in particular is simple enough to be independently useful or falsifiable by others working on time-aware recommendation, and we hope the preference-noise conjecture (Insight 2) is useful as a starting point for a more rigorous treatment of why frequency-domain modeling helps in recommendation, even if we were not able to fully substantiate it ourselves.

\bibliography{main}
\bibliographystyle{icml2024}
\end{document}